\documentclass[
  journal=pasa,
  manuscript=research-paper, %% or "review"
  year=2025,
  volume=YY,
]{cup-journal}

\usepackage{aas-macros}
\received {dd Mmm YYYY}
\revised  {dd Mmm YYYY}
\accepted {dd Mmm YYYY}
\published{22 September 202X}
\usepackage{amsmath, amssymb}
\usepackage[nopatch]{microtype}
\usepackage{booktabs}

\usepackage{multirow}
\usepackage{graphicx, float}	% Including figure files
\usepackage{subcaption} % for multiplot
\usepackage{parcolumns} % for multicolumn

\usepackage{hyperref}
\hypersetup{
  colorlinks   = true, 
  urlcolor     = blue,
  linkcolor    = blue, 
  citecolor    = blue
}

\newcommand{\pleunisinprep}{Pleunis et al. in prep}
\newcommand{\hoffmannetalinprep}{Hoffmann et al. in prep}

\newcommand{\DMISM}{DM$_{\rm MW,ISM}$}
\newcommand{\DMIGM}{DM$_{\rm cosmic}$}
\newcommand{\DMEG}{DM$_{\rm EG}$}
\newcommand{\DMG}{DM$_{\rm MW}$}
\newcommand{\DMhalo}{DM$_{\rm MW,halo}$}
\newcommand{\DMobs}{DM$_{\rm obs}$}
\newcommand{\DMhost}{DM$_{\rm host}$}

\newcommand{\DMunit}{pc\:cm$^{-3}$}

\newcommand{\sigmaHost}{$\sigma_{\rm host}$}
\newcommand{\muHost}{$\mu_{\rm host}$}

\newcommand{\nsfr}{$n$}
\newcommand{\Emax}{$E_{\mathrm{max}}$}
\newcommand{\Emin}{$E_{\mathrm{min}}$}

\newcommand{\FAST}{FAST}

\newcommand{\flyseye}{CRAFT Fly's Eye}
\newcommand{\icslow}{CRAFT/ICS 900 MHz}
\newcommand{\icsmid}{CRAFT/ICS 1.3 GHz}
\newcommand{\icshigh}{CRAFT/ICS 1.6 GHz}
\newcommand{\parkes}{Parkes/Mb}

\newcommand{\HOunit}{km\,$\rm s^{-1}$\,$\rm Mpc^{-1}$}

\newcommand{\Result}{$68^{+27}_{-24}$\,\DMunit{}}

\newcommand{\edit}[1]{\textcolor{black}{#1}}

\title{I can see your halo: Constraining the Milky Way halo DM with FRB population studies}

\author{J.~Hoffmann}
\affiliation{International Centre for Radio Astronomy Research, Curtin University, Bentley, WA 6102, Australia}
\email[J. Hoffmann]{jordan.hoffmann@postgrad.curtin.edu.au}

\author{C.~W.~James}
\affiliation{International Centre for Radio Astronomy Research, Curtin University, Bentley, WA 6102, Australia}
\author{J.~X.~Prochaska}
\affiliation{Department of Astronomy and Astrophysics, University of California, Santa Cruz, CA 95064, USA}
\alsoaffiliation{Kavli Institute for the Physics and Mathematics of the Universe, 5-1-5 Kashiwanoha, Kashiwa 277-8583, Japan}
\alsoaffiliation{Division of Science, National Astronomical Observatory of Japan, 2-21-1 Osawa, Mitaka, Tokyo 181-8588, Japan}

\author{M.~Glowacki}
\affiliation{Institute for Astronomy, University of Edinburgh, Royal Observatory, Edinburgh, EH9 3HJ, United Kingdom}
\alsoaffiliation{International Centre for Radio Astronomy Research, Curtin University, Bentley, WA 6102, Australia}
\alsoaffiliation{Inter-University Institute for Data Intensive Astronomy, Department of Astronomy, University of Cape Town, Cape Town, South Africa}

 %% First letter not capped

\begin{document}

\begin{abstract}
Fast radio bursts (FRBs) probe the electron column density along the line of sight and hence can be used to probe foreground structures. One such structure is the Galactic halo. In this work, we use a total of 98 high Galactic latitude ($|b| > 20^\circ$) FRBs detected by ASKAP, Parkes, DSA and FAST with 32 associated redshifts to constrain the dispersion measure (DM) contribution from the Galactic halo. We simultaneously fit unknown FRB population parameters, which show correlations with the Galactic halo but are not completely degenerate. We primarily use an isotropic model for the halo, but find no evidence favouring \edit{a particular halo model}. We find \DMhalo{}=\Result{}, which is in agreement with other results within the literature. Previous constraints on \DMhalo{} with FRBs have used a few, low-DM FRBs. However, this is highly subject to fluctuations between different lines of sight, and hence using a larger number of sightlines as we do is more likely to be representative of the true average contribution. Nevertheless, we show that individual FRBs can still skew the data significantly and hence will be important in the future for more precise results.
\end{abstract}

%==============================================================================
\section{Introduction} \label{sec:introduction}
%==============================================================================
Fast radio bursts (FRBs) are short, highly energetic radio signals originating at cosmological distances. The relatively young field of FRBs has rapidly developed since their discovery in 2007 \citep{Lorimer2007}, now with thousands of sources detected \citep[e.g.][]{CHIME2021} and of order a hundred localised to host galaxies \edit{\citep[e.g.][]{Rajwade2022, Shannon2024, Sharma2024, CHIME2025, PastorMarazuela2025}}. Current research in this field is broadly focused in two directions: understanding the origins of these bursts, and using their observables to study the cosmology of our Universe. In this work, we focus on population studies of FRBs, which hold information regarding both facets, and place particular emphasis on probing the baryon content of the Milky Way halo (\DMhalo{}).

FRBs have found an incredible niche in cosmological science due to their unique ability to directly measure the baryon content of the Universe \edit{\citep[e.g][]{Thornton2013, Macquart2020, Glowacki2024a}}. As electromagnetic radiation passes through ionised gas it experiences a frequency-dependent retardation. This is easily measured as an observable of the burst known as the dispersion measure (DM), which is directly proportional to the integrated electron column density along the line of sight. As baryons trace the electron distribution of the Universe, this allows FRBs to map out the \edit{complete} cosmological baryon distribution, which is otherwise unobservable \citep{Macquart2020, Khrykin2024, Connor2025}.

Most FRB studies focus on understanding the contribution of intergalactic and circumgalactic media (IGM and CGM) to the cosmological DM budget. However, FRBs are also one of the few direct probes of the Milky Way halo \edit{\citep{Prochaska2019a, Keating2020, Platts2020, Cook2023, Ravi2023}}. Measuring the baryonic content and distribution of the Galactic halo is notoriously difficult due to the diffuse nature of the plasma contained within it, existing in a range of phases with temperatures $T \sim 10^4 - 10^7$\:K. Typically, cool gas ($T<10^4$\:K) is clumpy and observable through 21\:cm HI emission, warm and warm-hot ($10^4 < T < 10^6$\:K) gas is observable through H$\alpha$ emission lines and absorption lines from quasi-stellar background objects, and hot ($T>10^6$\:K) gas is observable through X-ray emission and absorption lines of highly ionized metals, e.g. O$^{+7}$.

Current literature suggests that gas in lower temperature phases ($T<10^6$\:K) does not contribute significantly ($<15\%$) to the total Galactic baryon budget \citep[e.g.][]{Das2021} and hence X-ray emission and absorption analyses are currently the best estimators for the total electron content. X-ray emission spectroscopy is a direct observation of the gas in the Milky Way and so samples large areas of the sky without the need for a background source \edit{\citep[e.g.][]{Nakashima2018}}. However, it is biased towards denser gas and higher emissivity, meaning it only probes a subset of the total gas in the halo.

Conversely, X-ray absorption using background quasars does not have a density or emissivity bias. However, this only samples a small area in the sky due to the necessity for an X-ray bright quasar as a back-light \edit{\citep[e.g.][]{Fang2015}}.  Furthermore, these rely on metal lines, which implies an inherent degeneracy between the assumed or estimated metallicity and the total mass.

DM is a powerful observable because it is a direct measurement of all ionised gas along the line of sight, regardless of the quantity, phase and metallicity of the gas.  For a predominantly ionised gas, as anticipated for the halo of our Galaxy, DM provides an estimate of the baryonic mass along a given sightline. Additionally, as the number of FRB sightlines across the sky continues to rapidly rise, one may probe directional dependences. However, a significant limitation is the inability to disentangle the distribution of electrons along a given sightline, \edit{although such a degeneracy could be broken by other measures such as scattering in the future \citep{Cordes2022, MasRibas2025}}. The measured DM of an FRB has contributions from plasma in the local environment of the progenitor, the interstellar medium (ISM) and the halo of the host galaxy (\DMhost{}), cosmological gas in the \edit{IGM and the CGM} of intervening halos (\DMIGM{}), the Milky Way halo (\DMhalo{}) and the Milky Way ISM (\DMISM{}). DM is a measure of the integrated column density of electrons and, as such, cannot determine the relative contributions from each component. Therefore, meaningful constraints on the Galactic halo dispersion measure \DMhalo{} require specially designed 
experiments. 

One of these is to analyse the distribution of observed FRB DMs and estimate the minimum value after accounting for the Galactic ISM contribution. This minimum DM provides an upper limit to \DMhalo{} subject to the unknown host and IGM contributions. \citet{Platts2020} and \citet{Cook2023} used this technique to infer \DMhalo~$< 123$~\DMunit and $< 52-111$~\DMunit\ respectively. A variant of this approach is to restrict to FRBs associated at high confidence to a host galaxy with a measured redshift. In this case, one may estimate the cosmic dispersion measure \DMIGM\ and can even limit its contribution by restricting to very nearby (low redshift) FRBs. \citet{Ravi2023} took this approach with FRB~20220319D to infer \DMhalo~$< 47.3$~\DMunit.

These techniques are currently \edit{limited} to a few sightlines and are therefore subject to uncertainties in \DMISM\ along the individual sightlines, which directly translates to uncertainty in \DMhalo.  Furthermore, \DMhalo\ may be anisotropic \edit{\citep{Nakashima2018, Kaaret2020, Yamasaki2020, Das2021}}, and these works \edit{\citep{Platts2020, Cook2023, Ravi2023}}, which sample single or a few sightlines, are subjected to such halo fluctuations. As such, analyses of the lowest DM FRBs can inform us about the minimum densities within the Milky Way halo, but may or may not be indicative of a characteristic contribution. \edit{As the Canadian Hydrogen Intensity Mapping Experiment (CHIME) now have operational outriggers \citep{CHIMEOutriggers2025} to localise their detections, we expect the number of localised, nearby FRBs to increase substantially and as we fill out this parameter space, such a sample will make these methods significantly more robust in the future.} 
%must come from nearby FRBs in which \DMIGM{} (and thereby the corresponding scatter in \DMIGM{}) is small \citep[e.g.][]{Ravi2023, Cook2023} or 

Yet another method, and the one considered here, is to constrain gas profiles using a large number of FRBs with precise redshifts \citep{Mcquinn2014}. Such population studies require a large number of FRBs to obtain such estimates. In this work, we place constraints on \DMhalo{} using a large sample of 98 FRBs and 32 associated redshifts while simultaneously fitting other unknown population parameters. Our approach is statistical in nature and, in principle, allows for modelling the halo with an inhomogeneous density distribution.

In Section \ref{sec:methods_and_data}, we give an overview of the data that we use and our choices of data cuts. \edit{We also justify model choices and outline changes to the analysis techniques presented in \citet{james2022} and \citet{Hoffmann2025}}. We present our results in Section \ref{sec:results} and conclude in Section \ref{sec:conclusion}.

%==============================================================================
\section{Methods and data} \label{sec:methods_and_data}
%==============================================================================
This work is a continuation of previous publications using the \texttt{zDM} code base \citep{james2022}. The specific implementations are identical to those described in \citet{james2022b} and \citet{Hoffmann2025} except where otherwise specified. \edit{We model \DMhost{} as a log-normal distribution with mean \muHost{} and standard deviation \sigmaHost{} as free parameters; the cosmic evolution of FRB progenitors is modeled by the cosmic star formation rate to some power \nsfr{}; the spectral dependence is denoted by $\alpha$ where we use a `rate-interpretation' \citep{james2022} which is further discused in Section \ref{sec:parameters}; we model the luminosity function as an upper-incomplete Gamma function with slope $\gamma$, exponential cutoff beginning at \Emax{} and a hard-cutoff at some minimum energy \Emin{}. We additionally assume a flat $\Lambda$CDM model for our Universe and utilise measurements from Planck. We model telescope and algorithm baises \citep[e.g.][]{Hoffmann2024} for each telescope, and construct a DM model as in Equation \ref{eq:DM}. Using this model, we conduct a full likelihood analysis in a Markov-Chain Monte-Carlo (MCMC) framework to simultaneously constrain each of the unknown parameters.}

\subsection{Data}
We use the same base data as \citet{Hoffmann2025} with the addition of the data described in Table \ref{tab:FRBs}. This data includes FRBs from blind searches using the Australian SKA Pathfinder (ASKAP) telescope as part of the Commensal Real-time ASKAP Fast Transients (CRAFT) survey in the Fly's Eye and incoherent sum (ICS) modes, the Parkes Murriyang telescope, the Five-hundred meter Apperture Spherical radio Telescope (FAST) and the Deep Synoptic Array (DSA). The CRAFT/ICS survey is modelled in 3 different observing bands. 

We apply the same data selection cuts as per \citet{Hoffmann2025}, where we exclude redshift information for FRBs with extragalactic DM (\DMEG{}$=\rm{DM_{total}} -$\DMISM{}$-\langle$\DMhalo{}$\rangle$) exceeding that of the lowest, hostless FRB \edit{in the given survey. This cut is applied to the DSA dataset as it is unclear why each FRB does not have an associated redshift, and hence we cannot include them without bias.} We do not need to apply the same cut to FRBs detected with ASKAP, as the reason they do not have an associated redshift is due to the host galaxies not being followed up yet, and hence, there is no bias against high-redshift, low-DM objects. For the same reason, we now also include the redshift FRB 20171020A, which we previously excluded. 

We apply an additional Galactic latitude cut of $|b| > 20^\circ$ as low-latitude FRBs have unreliable estimates for \DMISM{} which can bias our results (see Section \ref{sec:uncertainties} for a further explanation). Given these cuts, we utilise a total of 98 FRBs and associated spectroscopic redshifts for 32 of them. A breakdown of the number of FRBs used in each survey is given in Table \ref{table:surveys}. This data includes 6 more FRBs detected in 2024 during the \icslow{} survey \citep{Shannon2024} and the redshift of FRB 20171020A detected during the \flyseye{} survey, which was previously excluded \citep{Lee-Waddell2023}. The details for these FRBs are presented in Table \ref{tab:FRBs}.

Although more FRBs and localisations from DSA and MeerKAT have been released, a complete sample has not been made available, which can introduce bias in our results. In particular, it is easier to detect host galaxies that are at a lower redshift, and so we would preferentially obtain low-redshift host galaxies for a given DM. Hence, until a full sample of all detected FRBs is made available, these data sets cannot be included in our analysis without bias. Additionally, providing the reasons as to why FRBs have not been associated with a host galaxy can allow more redshifts to be used, as the bias is only present if the redshift is absent due to a particularly distant or faint host galaxy. Thus, we strongly encourage authors to publish such information in future publications, and we are committed to providing it.

\edit{CHIME has the largest number of detected FRBs to date \citep{CHIME2021}, with an increasing number being accurately localised to host galaxies \citep{CHIMEbaseband2024, CHIME2025}. Due to its large field of view but relatively lower sensitivity, many of these FRBs are in the local Universe, including FRB 20200120E \citep{Bhardwaj2021} located in M81 and FRB 20250316A \citep{CHIMElocal2025} at a distance of 40\, Mpc. In principle, these FRBs are the most constraining on \DMhalo{} and hence would increase our constraining power significantly. However, CHIME has a unique bias towards detecting repeaters, which must be explicitly modelled \citep{James2023}. As this model is currently not fully functional within \texttt{zDM}, we cannot include CHIME FRBs without bias, and so leave this analysis to future work (\hoffmannetalinprep{}).}

\begin{table}
\begin{center}
\caption{The total number of FRBs and the corresponding number of redshifts that we use in our analysis from each survey after the given cuts.}
\label{table:surveys}
\begin{tabular}{lcc}
% \hline
% Survey & Number of FRBs & Number of redshifts \\
% \hline 
% \parkes{} & 12 & 0 \\
% \flyseye{} & 25 & 1 \\
% \icslow{} & 21 & 14 \\
% \icsmid{} & 19 & 13 \\
% \icshigh{} & 3 & 3 \\
% DSA & 25 & 3 \\
% \FAST & 9 & 0 \\
% \hline
% Total & 125 & 34 \\
% \hline
\hline
Survey & Number of FRBs & Number of redshifts \\
\hline 
\parkes{} & 16 & 0 \\
\flyseye{} & 25 & 1 \\
\icslow{} & 21 & 14 \\
\icsmid{} & 19 & 13 \\
\icshigh{} & 3 & 3 \\
DSA & 11 & 1 \\
\FAST & 3 & 0 \\
\hline
Total & 98 & 32 \\
\hline
\end{tabular}
\end{center}
\end{table} 

\begin{table}
\begin{center}
% \caption{Information used in addition to those in \citet{Hoffmann2024}. The FRBs listed from the \flyseye{} and DSA surveys were used in the previous analysis however the redshift information was not previously used for the given FRBs. FRBs which are below the $b < 20^\circ$ cutoff are not included. The redshift for FRB 20171020A was taken from \citet{Lee-Waddell2023} and the redshifts for all of the DSA FRBs were taken from \citet{Sharma2024}. The FRBs listed from the \icslow{} survey were not used at all in previous analyses and are further described in \citet{Shannon2024}.}
\caption{Information used in addition to those in \citet{Hoffmann2025}. FRBs that are below the Galactic latitude cut of $|b| < 20^\circ$ are not included. FRB 20171020A was used in previous analyses, however, the redshift was not previously utilised. The redshift was taken from \citet{Lee-Waddell2023}. The FRBs listed from the \icslow{} survey were not used at all in previous analyses and are further described in \citet{Shannon2024}.}
\label{tab:FRBs}
\begin{tabular}{cccccc}
\hline 
Name & \DMobs{} & \DMISM{} & $\nu$ & S/N & $z$ \\ %& $P_{\mathrm{host}}$ \\
& (\DMunit{}) & (\DMunit{}) & MHz & & \\ 
\hline 
\multicolumn{6}{c}{\flyseye{}} \\
\hline
% 20220501C & 449.5 & 30.6 & 863.5 & 16.1 & 0.381 \\
20171020A* & 114.1 & 38.4 & 1196 & 19.5 & 0.00867 \\
\hline
\multicolumn{6}{c}{\icslow{}} \\
\hline
20240201A & 374.5 & 38 & 920.5 & 13.9 & 0.043 \\ 
20240208A & 260.2 & 98 & 863.5 & 12.1 & - \\
20240210A & 283.7 & 31 & 863.5 & 11.6 & 0.024 \\
20240304A & 652.6 & 30 & 832.5 & 12.3 & - \\
20240310A & 601.8 & 36 & 902.5 & 19.1 & 0.127 \\
20240318A & 256.4 & 37 & 902.5 & 13.2 & - \\
% \hline
% \multicolumn{6}{c}{DSA} \\
% \hline
% 20220121B & 313.421 & 79.99 & 1405 & 9.38 & -  \\
% 20220204A & 612.584 & 52.58 & 1405 & 16.22 & 0.401 \\
% 20220208A & 440.73 & 88.37 & 1405 & 13.77 & 0.351 \\
% 20220307B & 499.328 & 120.02 & 1405 & 11.91 & 0.248 \\
% 20220310F & 462.657 & 45.46 & 1405 & 68.41 & 0.478 \\
% 20220330D & 467.788 & 38.42 & 1405 & 12.94 & 0.371 \\
% 20220418A & 624.124 & 36.35 & 1405 & 10.88 & 0.622 \\
% 20220424E & 863.932 & 132.80 & 1405 & 9.41 & - \\
% 20220506D & 396.651 & 82.85 & 1405 & 48.92 & 0.300 \\
% 20220726A & 686.232 & 79.72 & 1405 & 12.72 & 0.362 \\
% 20220801A & 413.416 & 101.63 & 1405 & 9.25 & - \\
% 20220825A & 649.893 & 77.31 & 1405 & 15.06 & 0.241 \\
% 20220831A & 1146.14 & 105.95 & 1405 & 19.19 & 0.262 \\
% 20220914A & 630.703 & 54.39 & 1405 & 9.64 & 0.114 \\
% 20220920A & 314.977 & 39.64 & 1405 & 14.35 & 0.158 \\
% 20220926A & 441.984 & 104.28 & 1405 & 10.26 & - \\
% 20221002A & 319.951 & 51.47 & 1405 & 8.50 & - \\
% 20221012A & 440.358 & 54.06 & 1405 & 9.41 & 0.285 \\
% 20221027A & 452.723 & 47.13 & 1405 & 12.13 & 0.542 \\
% 20221029A & 1391.746 & 43.13 & 1405 & 12.06 & 0.975 \\
% 20221101B & 491.554 & 116.47 & 1405 & 10.12 & 0.240 \\
% 20221101A & 1475.53 & 79.69 & 1405 & 14.97 & - \\
\hline
\end{tabular} 
\end{center} 
\end{table} 

\subsection{Parameters} \label{sec:parameters}
We adopt the same base parameters and priors for FRB population modelling as \citet{Hoffmann2025}, except for our prior on the spectral index $\alpha$. We have changed this prior because we exclude information about the number of events detected in a given survey, $P(N)$. While $P(N)$ does contain useful information, the difficulty in characterising the observational time spent on the sky has posed many challenges, and we consider it an unreliable quantity. Even amongst surveys conducted with ASKAP, the rates that have been obtained have been inconsistent \citep{Shannon2024, Hoffmann2025}. Without $P(N)$, we cannot obtain good constraints on $\alpha$, and so we place a restrictive prior on it. Results previously obtained from population studies have had large ($\sim100\%$) uncertainties \citep[e.g.][]{james2022, shin2023, Hoffmann2025}, and thus, values of $\alpha$ obtained through direct analysis of FRB spectra are expected to be more reliable. Hence, we defer to priors from such studies. 

\edit{When considering other studies, it is important to understand the interpretation for $\alpha$ that they imply. Our model broadly modifies the detection rate of FRBs at a given frequency ($\nu$) by $\nu^{\alpha}$. As discussed in \citet{james2022}, for a negative $\alpha$, this can be interpreted as:
\begin{enumerate}
    \item Assuming all bursts are broadband, then lower frequencies contain more energy (spectral index interpretation; $\alpha_{\mathrm{SI}}$).
    \item Assuming bursts are narrowband, then there is a greater number of bursts at lower frequencies but with similar energies (rate interpretation; $\alpha_{\mathrm{R}}$).
\end{enumerate}
Previously, we have used a rate interpretation due to its lower computational requirements, however, most measurements in the literature use a spectral index interpretation. We also note that the choice of model primarily affects \nsfr{} and $\alpha$ itself, while not having large impacts on the other parameters, including those coming from the DM budget \citep{Hoffmann2025}.} 

\edit{The first measurement of $\alpha$ was taken by \citet{Macquart2019} who examined spectra from 23 FRBs detected by ASKAP as part of the Commensal Real-time ASKAP Fast Transients (CRAFT) survey and found $\alpha_{\mathrm{SI}} = -1.5^{+0.2}_{-0.3}$, corresponding to $\alpha_{\mathrm{R}} = -0.63 \pm 0.3$ \citep{james2022}. More recently, \citet{Shannon2024} examined the frequency dependence of the FRB event rate and noted that we have poor constraints of $\alpha_{\mathrm{SI}} = -0.3^{+1.4}_{-1.6}$ for the extended CRAFT incoherent sum (ICS) observations. Similar studies using 62 broadband FRBs detected by the CHIME found $\alpha_{\mathrm{SI}} = -0.98 \pm 0.05$ (\pleunisinprep), but \citet{Cui2025} measure a steeper value of $\alpha = -2.29 \pm 0.29$ with the CHIME sample when not restricting their analysis to broadband pulses (and hence being closer to $\alpha_{\mathrm{R}}$). As such, while using the rate interpretation, we apply a uniform prior on $\alpha_{\mathrm{R}}$ of -0.5 to -2.5 and do not utilise $P(N)$ any longer.}

All other priors are unrestrictive except for the Hubble constant, $H_0$, on which we impose a uniform prior of 66.9 to 73.08 \HOunit: a window spanning 1\,$\sigma$ either side of the best estimates from \citet{SH0ES2021} and the \citet{Planck2018}. 

Additionally, in this work, we are primarily interested in finding constraints on \DMhalo{}. As such, we allow the mean value of \DMhalo{} to vary as a free parameter while imposing a uniform linear prior from 0 to 100\:\DMunit{}. We find that there is no significant difference when using log or linear priors on \DMhalo{}, and as the summation is fundamentally linear (see Equation \ref{eq:DM}), we choose to use a linear uniform prior. \edit{Alternatively, \muHost{} and \sigmaHost{} use log-uniform priors to describe the log-normal distribution of \DMhost{}.}

\begin{table}
\begin{center}
\caption{Limits on the uniform priors used. The parameters are as follows: \nsfr{} gives the correlation with the cosmic SFR history; $\alpha$ is the slope of the spectral dependence; $\mu_{\mathrm{host}}$ and $\sigma_{\mathrm{host}}$ are the mean and standard deviation of the assumed log-normal distribution of host galaxy DMs; \Emax{} notes the exponential cutoff of the luminosity function (modelled as a Gamma function); \Emin{} is a hard cutoff for the lowest FRB energy; $\gamma$ is the slope of the luminosity function; and $H_0$ is the Hubble constant. \DMhalo{} is in units of \DMunit{}. The host parameters $\mu_{\mathrm{host}}$ and $\sigma_{\mathrm{host}}$ are in units of \DMunit{} in log space, \Emax{} and \Emin{} are in units of ergs and $H_0$ is in units of km$\:$s$^{-1}\:$Mpc$^{-1}$. The limits on $\alpha$ were informed by existing measurements in the literature. The limits on \Emax{} and \Emin{} were chosen as the distributions are uniform on the extrema of these ranges. The limits on $H_0$ represent a 1\,$\sigma$ interval around the \citet{Planck2018} and \citet{SH0ES2021} results.}
\label{table:params}
\begin{tabular}{lcc}
\hline
Parameter & Prior Min & Prior Max \\
\hline 
\DMhalo{} & 0.0 & 300 \\
\nsfr{} & -2.0 & 6.0 \\
$\alpha$ & -2.5 & -0.5 \\
$\mu_{\mathrm{host}}$ & 1.0 & 3.0 \\
$\sigma_{\mathrm{host}}$ & 0.1 & 1.5 \\
log$_{10}$(\Emax{}) & 40.35 & 45.0 \\
log$_{10}$(\Emin{}) & 36.0 & 40.35 \\
$\gamma$ & -3.0 & 0.0 \\
$H_0$ & 66.9 & 74.08 \\
\hline
\end{tabular}
\end{center}
\end{table} 

\subsection{Degeneracy of \DMhalo{} and \DMhost{}} \label{sec:degeneracy}
We model the total DM of each FRB as
\begin{equation}
    \mathrm{DM_{total} = DM_{MW,ISM} + DM_{MW,halo} + DM_{cosmic}} + \frac{\mathrm{DM_{host}}}{1+z}, \label{eq:DM}
\end{equation}
with contributions from the Milky Way (ISM), the Milky Way halo, cosmic ionised gas and the host galaxy, respectively. Of these, \DMhalo{} and \DMhost{} are poorly constrained, and hence we fit for them in our analysis. \edit{We model the distribution of \DMhalo{} values as a linear-normal distribution with a variable mean and a fixed standard deviation of 15\,\DMunit{} as discussed in Section \ref{sec:uncertainties}, and \DMhost{} as a log-normal distribution with mean (\muHost{}) and standard deviation (\sigmaHost{}) as free parameters}. We choose a log-normal distribution for the host galaxy, as FRBs can be found far outside their galaxies' ISMs \citep[e.g.][]{Kirsten2022}, and hence \DMhost{} has a hard cutoff at 0\,\DMunit{} but no effective upper limit allowing an upwards tail. On the contrary, the Earth is embedded well within the Galactic halo, and so we expect relatively consistent values of \DMhalo{}. \edit{Ultimately, when experimenting with linear-normal and log-normal distributions of \DMhalo{}, our results did not differ significantly and so this choice is not significant.}

Although \DMhalo{} and \DMhost{} have similar effects, the difference in the chosen functional form of their distributions and the $(1+z)^{-1}$ weighting on \DMhost{} allows this degeneracy to be broken, given a sufficient number of FRBs. Of course, if the Universe conspires FRB host galaxies to have \DMhost\ with a redshift dependence that approaches \DMhost~$\sim (1+z)$, then the degeneracy between these two components will be more difficult to resolve.

% From our DM model, we then obtain
% \begin{eqnarray}
%     \mathrm{DM_{MW,halo}} &=& \mathrm{DM_{total} - DM_{MW,ISM} - DM_{cosmic}} \nonumber \\
%     && - \,\frac{\mathrm{DM_{host}} }{(1+z)} \label{eq:DM}
% \end{eqnarray}
% assuming \DMhost{} is independent of the other DM contributions and where \DMhost{} is log-normally distributed with a mean of \muHost{} and a standard deviation of \sigmaHost{}. We therefore expect a degeneracy between \DMhalo{} and \muHost{}.

% We also note that the specific computational implementation differs for \DMhalo{} and \DMhost{}. For \DMhalo{}, we apply a constant shift of \muHalo to all FRBs uniformly before any analysis is completed. Then, when calculating the likelihood of an FRB being detected at the given DM, we integrate over a Gaussian uncertainty in possible values \citep{Hoffmann2024}. For \DMhost{}, we apply a smearing factor to the $P(z,\mathrm{DM})$ grid \citep{james2022b}.

\subsection{\DMG{} uncertainties} \label{sec:uncertainties}
We previously demonstrated that large uncertainties in \DMISM{} introduce significant uncertainties in cosmological parameters such as $H_0$ \citep{Hoffmann2025}. As such, it is important to minimise these errors. In particular, sightlines passing through the Galactic plane have significantly more variation than those at high Galactic latitudes. While we do implement a percentage uncertainty on \DMISM{} to account for this, this causes underfluctuations to be considered more precise and overfluctuations less precise, resulting in a bias towards lower \DMISM{} values. Additionally, the fluctuations at low galactic latitude are significantly greater (even when considering percentage uncertainties) than those at high latitude, and hence applying a uniform uncertainty is not sensible. Thus, in this analysis, we exclude all FRBs detected below a Galactic latitude of $|b| < 20^\circ$. \edit{Above this latitude, uncertainties in Galactic DM contributions are minimised due to smoother gas distributions and the rarity of discrete structures \citep{Ocker2020, Ocker2024}. As such, we reduce the 50\% uncertainty on \DMISM{} estimated from \citet{Schnitzeler2012} to 20\%, but otherwise implement this uncertainty using the same method as \citet{Hoffmann2025}.}

We also note that little is known about \DMhalo{}, particularly regarding its distribution and homogeneity. While we do fit for the average value of \DMhalo{}, we do not account for fluctuations along different lines of sight (nor asymmetry with Galactic latitude or any other geometric parameter; see below). However, X-ray measurements have observed such variations, with \citet{Das2021} combining X-ray absorption and emission measurements obtaining a mean \DMhalo{} of $64^{+20}_{-23}$\,\DMunit{}. \citet{Yamasaki2020} note that there is a greater contribution from the halo in the direction of the Galactic disk, and thus by excluding these FRBs we expect to decrease the range of \DMhalo{} values which we probe. \edit{Thus, to account for possible variations in \DMhalo{} from differing sight-lines, we model the distribution of \DMhalo{} values as a normal distribution with a standard deviation of $15$\,\DMunit{} and fit for the mean.} Currently, we implement this as a linear uncertainty. However, \citet{Das2021} argue that the distribution of Galactic DM values has a positive skew even in log-space. While this choice is somewhat arbitrary and we do not expect it to have any significant impact on our current results, changing to a Gaussian uncertainty in log-space is a consideration for future analyses.

Although there have been models suggesting that the Galactic halo is not isotropic, most models include an additional term towards the direction of the Galactic disk \citep[e.g.][]{Yamasaki2020}. As we exclude FRBs at a low Galactic latitude, we continue with our assumption that the Galactic halo is approximately homogeneous in our analysis. A comparison of different models is explored in Section \ref{sec:models}.

\edit{In general, our estimates of \DMG{} are clearly not exact and so we implement Gaussian uncertainties on \DMG{} using the same method as \citet{Hoffmann2025}}. However, the choice of 20\% uncertainty on \DMISM{} and 15\,\DMunit{} uncertainty on \DMhalo{}, as well as modelling both as a Gaussian uncertainty, is somewhat arbitrary. In the future, when we have more localised FRBs, it may be possible to even constrain these uncertainties as free parameters.

\subsection{Fixing S/N calculation} \label{sec:S/N}

The probability, $p_s$, of detecting an FRB with signal-to-noise ratio S/N a factor $s$ above threshold S/N$_{\rm th}$ (i.e., ${\rm S/N} = s {\rm S/N}_{\rm th}$) in the range $s$ to $s + ds$, given that an FRB has already been detected, is given by
\begin{eqnarray}
\frac{dp_s}{ds} & = & \frac{L(s E_{\rm th}) \frac{dE}{ds}}{\int_{E_{\rm th}}^{\infty} L(E) dE}, \label{eq:ps}
\end{eqnarray}
where $L(E)$ is the FRB `luminosity' function (treated here as a Schechter function of energy $E$ in ergs assuming an emission bandwidth of $1$\,GHz), and $E_{\rm th}$ is the energy threshold corresponding to S/N$_{\rm th}$, and depending on FRB properties such as DM, $z$, position in the telescope beam $B$, and total effective FRB width $w_{\rm eff}$ \citep{james2022}. For whichever of these latter variables are unknown --- or otherwise their values for each FRB are not given --- the \texttt{zDM} code calculates the relative probability of each, and weights the final likelihood ${\mathcal L}_s$ as (in the case of an unlocalised FRB with known Galactic DM)
\begin{eqnarray}
{\mathcal L}_s & = & \int p_z(z) \int p_B(B) \int p_w(w_{\rm eff}) \frac{dp_s}{ds} dz \, dB \,dw_{\rm eff}, \label{eq:correct}
\end{eqnarray}
where for simplicity, the dependencies of probability distributions have been omitted.

However, in prior versions of the code, the integrations in Equation\ \ref{eq:correct} were applied seperately to the numerator and denominator of Equation~\ref{eq:ps}, producing
\begin{eqnarray}
{\mathcal L}_s  =  \frac{\int p_z(z) \int p_B(B) \int p_w(w_{\rm eff}) L(s E_{\rm th}) \frac{dE}{ds} dz \, dB \,dw_{\rm eff} }{\int p_z(z) \int p_B(B) \int p_w(w_{\rm eff}) \int_{E_{\rm th}}^{\infty} L(E) dE dz \, dB \,dw_{\rm eff} }. \label{eq:incorrect}
\end{eqnarray}
In consequence, regions of the $z$--$B$--$w_{\rm eff}$ parameter space with high probability $p_s$ given a detection (Equation~\ref{eq:ps}), but low probability of that detection (low $p_z$, $p_B$, and/or $p_{w}$ in Equation~\ref{eq:correct}), would be incorrectly up-weighted in probability, and vice-versa. This has now been fixed, so that the code implements Equation~\ref{eq:correct}.

%==============================================================================
\section{Results} \label{sec:results}
%==============================================================================
\begin{figure*}
\begin{center}
\includegraphics[width=\linewidth]{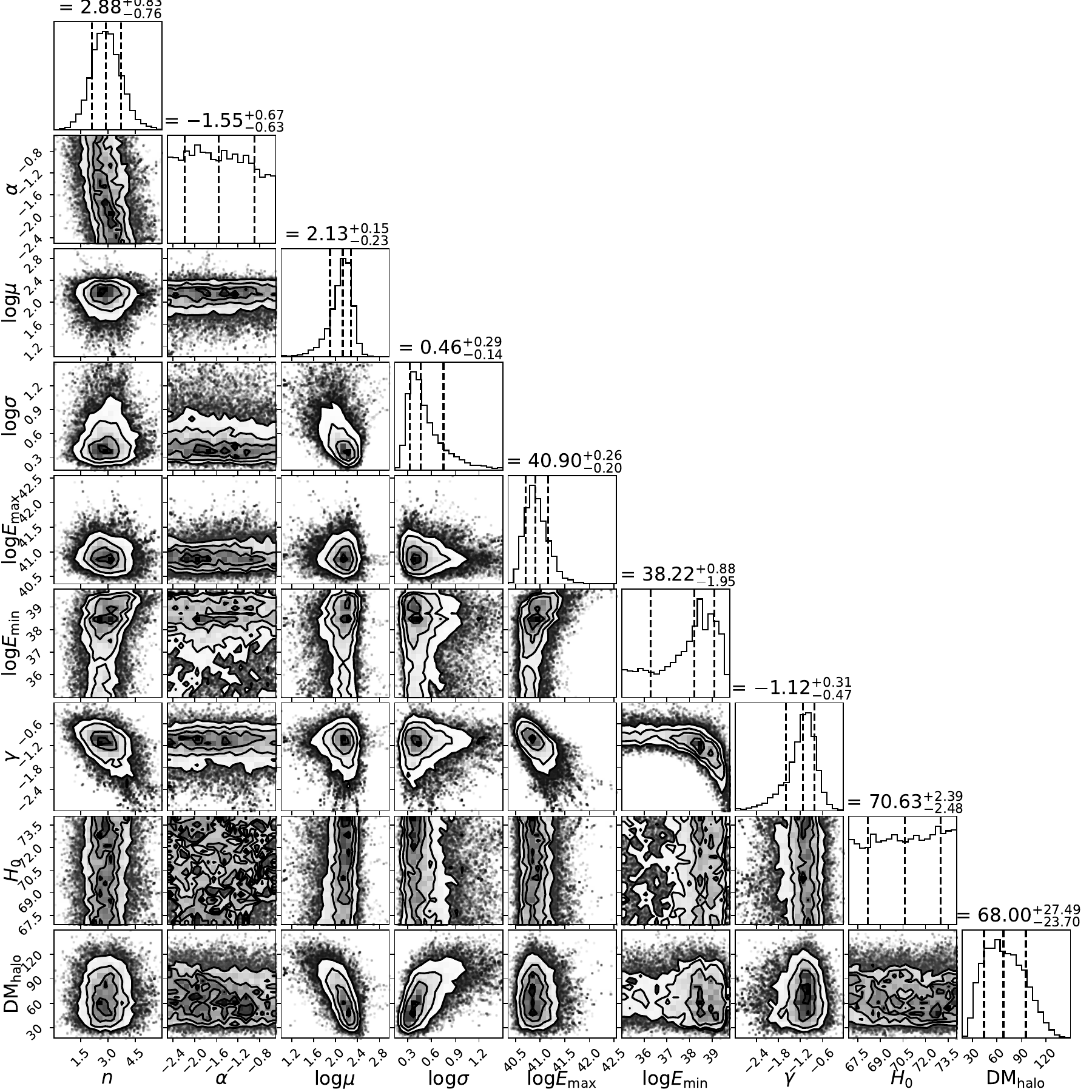}
\caption[]{Results from the MCMC analysis including FAST, DSA and CRAFT FRBs. The parameters are identical to those described in Table~\ref{table:params}.}
\label{fig:MCMCbase}
\end{center}
\vspace{-3ex}
\end{figure*}

We obtain a constraint on the mean value of \DMhalo{}=\Result. Figure \ref{fig:MCMCbase} shows our results for an MCMC analysis using 30 walkers, 3000 steps and a burn-in of 500. The values for most parameters do not differ significantly from the results of \citet{Hoffmann2025}, and hence we focus our discussion here on the results of \DMhalo{}. We note that our results for $\alpha$ and $H_0$ are a function of the priors, and we have no constraining power within the range set by the priors, as expected from Section \ref{sec:parameters}.

% The most notable difference is regarding \nsfr{} for which we no longer constrain an upper limit. It is clear that for a given value of $\alpha$, \nsfr{} has a strong lower limit but an unconstrained upper limit. The lower limit increases with $\alpha$ as well, which gives the impression that \nsfr{} prefers much larger values; however, the spectrum is relatively flat above the lower limit for each value of $\alpha$. Previously, we obtained an upper limit on \nsfr{} \citep{Hoffmann2025}, which is no longer present. We attribute the lack of an upper limit to corrections in how we calculate the probability of detecting each FRB at the observed S/N, $P_{\rm SNF}$ and not to the introduction of \DMhalo{} (see Section \ref{sec:S/N} for a more complete description).

\subsection{\DMhalo{} correlations}
\begin{figure}[H]
\begin{center}
\includegraphics[width=\linewidth]{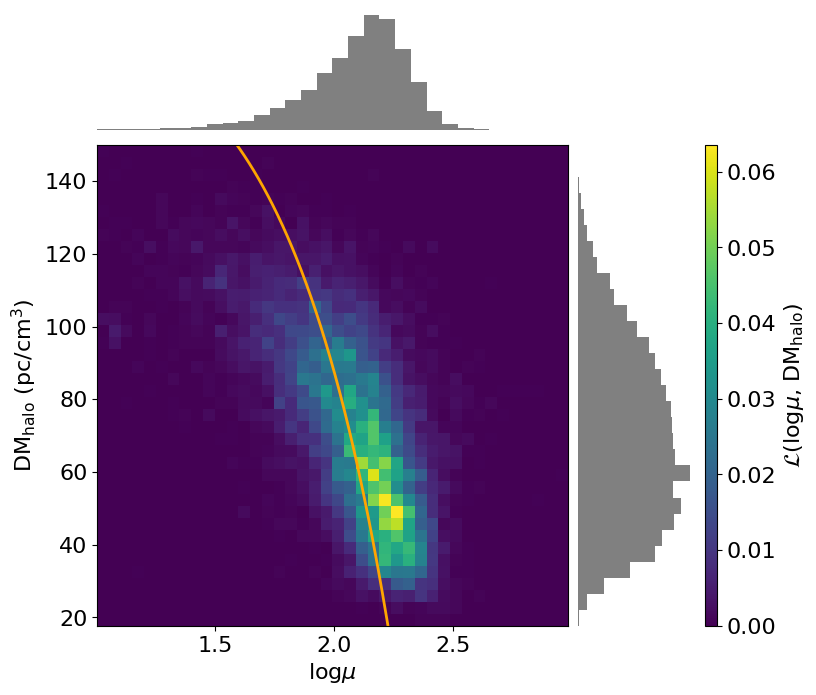}
\caption[]{Shown is the correlation between \DMhalo{} and \muHost{} from our MCMC analysis. Overplotted in orange is the expected degeneracy, calculated according to Equation \ref{eq:DM}.}
\label{fig:host_halo_corr}
\end{center}
\vspace{-3ex}
\end{figure}

Figure \ref{fig:host_halo_corr} shows the correlation between \DMhalo{} and the host galaxy parameter \muHost{}. The orange line shows the expected relationship from Equation \ref{eq:DM}. As expected, we observe an anti-correlation between these parameters, which is also evident between \DMhalo{} and \sigmaHost{}. While both \DMhalo{} and \muHost{} represent a constant, average contribution to each FRB in our analysis, the degeneracy between the two parameters is broken due to the $(1+z)^{-1}$ weighting factor on \DMhost{} caused by cosmic expansion and the difference in our chosen distributions (linear-normal and log-normal, respectively) as discussed in Section \ref{sec:degeneracy}.

We do not expect to see, and indeed do not see, strong correlations with other parameters. This confirms our previous assertions that our assumptions of \DMhalo{} would not have a large impact on any of our previous results with the exclusion of \DMhost{} parameters.

% \DMhalo{} also shows correlations with the luminosity function parameters \Emax{}, \Emin{} and $\gamma$. For FRBs with associated redshifts, we do not expect any correlation between \DMhalo{} and these parameters. Alternatively, for unlocalised FRBs, a higher \Emin{} suggests that bursts originate from further away and hence more of the DM budget is accounted for by \DMEG{}, giving a stronger upper limit on \DMhalo{} as is evident in our analysis. The correlations between \Emin{}, \Emax{} and $\gamma$ then cause all three parameters to be correlated with \DMhalo{}. No significant correlations are seen with $H_0$, \nsfr{} or $\alpha$.

%==============================================================================
\section{Discussion} \label{sec:discussion}
%==============================================================================
\subsection{Comparison with other results}
Previously, \citet{Cook2023} placed an upper limit on \DMhalo{} of $52-111$\,\DMunit{} using a sample of CHIME FRBs by observing a gap in DMs between \edit{Galactic pulsars and extragalactic FRBs.} This result is in good agreement with our own result of \DMhalo{}=\Result{}, however, our results are more constraining, which is to be expected when using a larger number of localised FRBs. \citet{Ravi2023} obtained a slightly lower upper limit of $47.3$\,\DMunit{} using a single nearby FRB. This result is in agreement with our own within $1\,\sigma$. The estimated \DMISM{} of this FRB exceeds the total DM of the source, which attests to the large uncertainties along this line-of-sight and hence we expect our result to be more robust.

In addition to measurements from FRBs, measurements from X-ray observations have been historically more prevalent. \citet{Prochaska2019a} estimated values of $50 \sim 80$\,\DMunit{} using X-ray absorption lines and \citet{Yamasaki2020} predicted a full range of $30 \sim 245$\,\DMunit{} with a mean of $43$\,\DMunit{} based on X-ray emission. \citet{Das2021} use a combination of emission and absorption lines from numerous elements and estimate \DMhalo{} in the range of $12-1749$\,\DMunit{} with a mean and a median of 161 and 64\,\DMunit{} respectively. X-ray absorption is preferentially biased against low densities, while DM probes all states of ionised gas. As such, it is unsurprising that results from X-ray absorption give lower predicted values. The results of \citet{Das2021} are in much closer agreement with our own, and these results have corrected for other states of gas that are not visible to X-ray probes.

These studies give a statistical distribution of direct measurements of \DMhalo{} and have shown a positive skew within this distribution \citep[see][]{Das2021}, causing the mean values to greatly exceed the median. \edit{We similarly see a positive skew in our results. However, our results of \DMhalo{} represent the distribution of the mean \DMhalo{} value and not an overall distribution of \DMhalo{} values given by \citet{Das2021}.}

\edit{The Large Magellanic Cloud (LMC) and Small Magellanic Cloud (SMC) also sit at a similar displacement from the Milky Way disk as the halo. As such, pulsars within the LMC and SMC can also inform us on the expected value for \DMhost{}. Using these pulsars, \citet{Anderson2010} estimate a low contribution of \DMhalo{}$=23$\,\DMunit{} which is 2$\sigma$ below our estimated value. However, the exact placement and extent of the halo is unclear, and hence whether the SMC and LMC lie within the halo or not is uncertain \citep[e.g.][]{Ravi2023}, meaning these measurements do not provide hard upper-limits on \DMhalo{}.}

\subsection{Impact of single low-DM FRBs} \label{sec:lowDM}
The primary advantage that our approach has over previous studies using FRBs to constrain \DMhalo{} is in using an ensemble of FRBs rather than single, nearby bursts. However, it is undeniable that these low-DM FRBs provide significant constraining power on \DMhalo{}. As such, we investigate the impact that a single FRB, such as FRB 20220319D, would have if we could get reliable \DMISM{} estimates.

Of the nearby FRBs, FRB 20220319D provides the strictest constraint, and was previously used to provide an upper limit on \DMhalo{} of 47.3\,\DMunit{} \citep{Ravi2023}. We have excluded this FRB from our analysis as it is at a low Galactic latitude ($b = 9.1^{\circ}$). In fact, this FRB is a primary example of why we exclude FRBs at such low Galactic latitudes as the predicted \DMISM{} contribution from both \verb|NE2001| and \verb|YMW16| \citep{Yao2017} exceeds the total DM of the FRB, and hence are necessarily incorrect. Even with more in-depth analysis using nearby pulsar DMs, the Galactic contribution is unreliable, as even adjacent sightlines through the Galactic plane have highly varying DM values \citep{Cordes2002, Das2021}. However, we still show the effect of such a nearby FRB if it were to be detected out of the Galactic plane with more reliable constraints.

\begin{figure}
\begin{center}
\includegraphics[width=\linewidth]{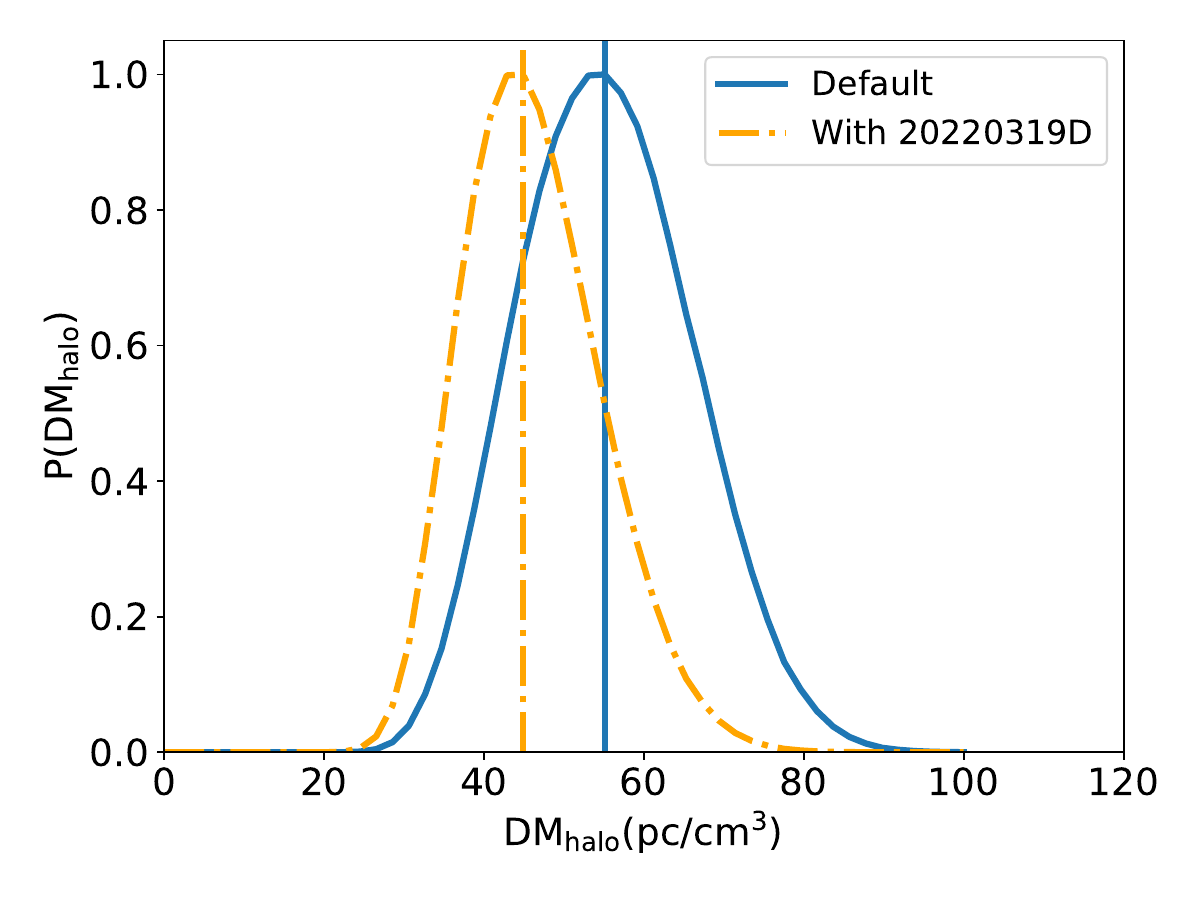}
\caption[]{A slice through \DMhalo{} when including or excluding FRB 20220319D. All other parameters are kept constant at their best fit values from Figure \ref{fig:MCMCbase}. When excluding FRB 20220319D we obtain \DMhalo{}=55\,\DMunit{} and when including it we obtain \DMhalo{}=45\,\DMunit{}.}
\label{fig:lowDM}
\end{center}
\vspace{-3ex}
\end{figure}

Figure \ref{fig:lowDM} shows our constraints on \DMhalo{} when keeping all other parameters fixed at their best-fit values as shown in Figure \ref{fig:MCMCbase}. The plot shows the difference of including this one FRB, with only a 20\% uncertainty on \DMISM{}. When excluding this FRB we obtain \DMhalo{}=55\,\DMunit{} and when including it this decreases to \DMhalo{}=45\,\DMunit{}. Thus, it is clear that these nearby FRBs do provide strong upper limits on \DMhalo{} as we would expect, and detections of such FRBs outside of the Galactic plane would greatly improve our results by providing stringent upper limits. 

% However, we still note that even with the inclusion of FRB 20220319D in our analysis, our prediction for the average \DMhalo{} value exceeds the upper limit obtained by \citet{Ravi2023} using this single FRB, and hence it is likely that the particular FRB passes through an underfluctuation in the Galactic gas distribution.

% \subsection{Biases from low latitude FRBs}
% \todo{Should I add a section including all FRBs and discuss the bias of FRBs passing through the Galactic centre to further justify their exclusion? (Decreases \DMhalo{} to $\sim$53)}

\subsection{Comparing halo models} \label{sec:models}
\begin{table}
\begin{center}
\caption{Median, mean ($\mu$) and standard deviation ($\sigma$) of the data from Figure \ref{fig:models}.}
\label{table:models}
\begin{tabular}{lccc}
\hline
Halo Method & Median$_{\Delta \mathrm{DM}}$ & $\mu_{\Delta \mathrm{DM}}$ & $\sigma_{\Delta \mathrm{DM}}$ \\
\hline 
Isotropic & -70 & -26 & 138 \\
\citet{Yamasaki2020} & -41 & 1.4 & 139 \\
\citet{Das2021} & 22 & 58 & 134 \\
\hline
\end{tabular}
\end{center}
\end{table} 

\begin{figure}
\begin{center}
\includegraphics[width=\linewidth]{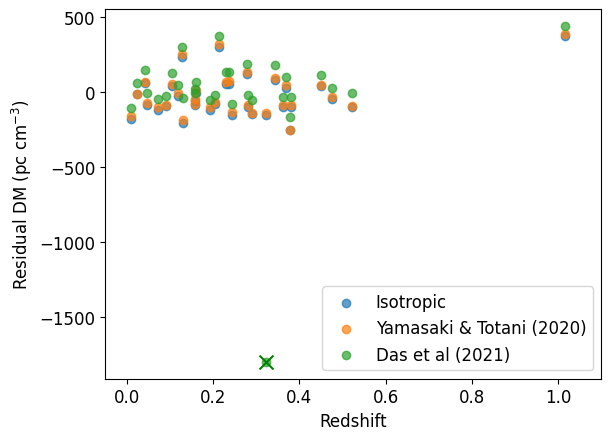}
\caption[]{Residual DMs ($\Delta \mathrm{DM}$) of the localised FRBs used in our analysis given different halo models. This represents the scatter around the Macquart relation. The three halo models considered were an isotropic halo, an empirical halo from X-ray observations \citep{Das2021} and an isotropic halo with an additional disk-like component \citep{Yamasaki2020}. The point from \citet{Das2021} at the bottom of the plot marked with a green cross is considered an outlier as the estimated \DMG{} is $1750^{+4550}_{-1370}$\,\DMunit{} and this estimation comes from a point 16.5 degrees away from the FRB position on the sky and hence is considered unreliable.}
\label{fig:models}
\end{center}
\vspace{-3ex}
\end{figure}

Our analysis assumes an isotropic halo, however, there has been evidence that there is structure within the Galactic halo \citep{Yamasaki2020, Das2021}. As such, we investigate three different halo models here by testing to see whether they reduce the scatter around the Macquart (mean $z-$DM) relation.

The first model is our assumed model of an isotropic halo, which takes on a single average value in all directions. The second is informed by X-ray observations and implements a smooth, isotropic, spherical halo superimposed with a disk-like component \citep{Yamasaki2020}. The third is an empirical model from \citet{Das2021} which uses 72 X-ray observations to map out the entire sky. This model is not smoothed and uses interpolation between nearby points to estimate the contribution for any given sight-line.

To calculate the scatter around the Macquart relation, we define $\Delta \rm DM$ as
\begin{eqnarray}
    \Delta \mathrm{DM} &\cong& \mathrm{DM_{tot} - DM_{NE2001} - DM_{MW,halo}} \nonumber \\
    && - \mathrm{DM_{Macquart}} - \frac{\langle \mathrm{DM_{host}} \rangle}{1+z}
\end{eqnarray}
for each FRB, where $\langle \mathrm{DM_{host}} \rangle =$10\^{}(\muHost{}+$\sigma_{\mathrm{host}}^2$/2). A plot of this scatter for each of the models is shown in Figure \ref{fig:models}. The corresponding median, mean, and standard deviation values are shown in Table \ref{table:models}.

The isotropic halo and the model of \citet{Yamasaki2020} perform very similarly with $\sigma_{\Delta \rm DM}$ values of 137 and 138\,\DMunit{}, respectively. The fundamental difference between the models is that \citet{Yamasaki2020} add an extra disk-like component, which is in line with the disk of the Milky Way. However, in our analysis, we place a Galactic latitude cut on FRBs with $|b| < 20^\circ$ and hence do not expect large differences between these models. The model of \citet{Das2021} also performs similarly with a scatter of 134\,\DMunit{}. However, there is an outlier amongst the sample which, when included, increases the scatter to 324\,\DMunit{}. The estimated \DMG{} for this FRB is $1750^{+4550}_{-1370}$\,\DMunit{} and this estimation comes from a point 16.5 degrees away from the FRB position on the sky. As such, we choose to exclude this point from our results.

\edit{Thus, we see no preference for any of the models of the Milky Way halo tested here, and would need more data to make definitive conclusions.}

%==============================================================================
\section{Conclusion} \label{sec:conclusion}
%==============================================================================
In this work, we use a large sample of 98 FRBs, alongside 32 associated redshifts at high ($b>20^{\circ}$) Galactic latitudes, to constrain the free electron column density in the Galactic halo. When fitting unknown FRB population parameters alongside \DMhalo{}, we obtain a value \edit{for the mean of} \DMhalo{}=\Result{}, which is in good agreement with existing literature using both FRBs and X-ray observations. 
%Our uncertainties are comparable to other estimates, and hence, with the rapidly growing number of localised FRBs, this will quickly become a competitive constraint.

%Our result is $\sim25$~\DMunit{} higher than the previous literature but within 1$\sigma$. Previous studies with FRBs used a small number of low-DM FRBs in the local Universe. As such, only a single line-of-sight was considered, which may not be representative of the average contribution. However, we also find that nearby FRBs do provide strong upper limits on \DMhalo{}, and hence, if such FRBs are detected away from the Galactic plane, they would significantly improve our constraints.

\edit{Our result is in good agreement with results from X-ray observations. It is higher than upper-limits provided by previous FRB studies, however, it is in agreement at the 1$\sigma$ level.} Previous studies with FRBs used a small number of low-DM FRBs in the local Universe. As such, only a single line-of-sight was considered, which may not be representative of the average contribution. However, we also find that nearby FRBs do provide strong upper limits on \DMhalo{}, and hence, if such FRBs are detected away from the Galactic plane, they would significantly improve our constraints.

Moving forward, we expect to obtain a large number of localised FRBs in the nearby Universe from large field-of-view surveys such as CHIME and DSA. This will allow us to more robustly constrain the average value \DMhalo{} but also to fit the level of fluctuations in \DMhalo{} which we currently select to be 15\,\DMunit{}. It may also become possible to measure a distribution of \DMhalo{} values and even probe directional dependence.

\begin{acknowledgement}
This work was performed on the OzSTAR national facility at Swinburne University of Technology. The OzSTAR programme receives funding in part from the Astronomy National Collaborative Research Infrastructure Strategy (NCRIS) allocation provided by the Australian Government. 

% This scientific work uses data obtained from Inyarrimanha Ilgari Bundara, the CSIRO Murchison Radio-astronomy Observatory. We acknowledge the Wajarri Yamaji as the Traditional Owners and native title holders of the Observatory site. CSIRO’s ASKAP radio telescope is part of the Australia Telescope National Facility. The operation of ASKAP is funded by the Australian Government with support from the National Collaborative Research Infrastructure Strategy. ASKAP uses the resources of the Pawsey Supercomputing Research Centre. The establishment of ASKAP, Inyarrimanha Ilgari Bundara, the CSIRO Murchison Radio-astronomy Observatory and the Pawsey Supercomputing Research Centre are initiatives of the Australian Government, with support from the Government of Western Australia and the Science and Industry Endowment Fund.

\end{acknowledgement}

\paragraph{Funding Statement}
This research was supported by an Australian Government Research Training Program (RTP) Scholarship.
CWJ and MG acknowledge support through Australian Research Council (ARC) Discovery Project (DP) DP210102103. MG is also supported by the UK STFC Grant ST/Y001117/1. MG acknowledges support from the Inter-University Institute for Data Intensive Astronomy (IDIA). IDIA is a partnership of the University of Cape Town, the University of Pretoria and the University of the Western Cape. For the purpose of open access, the author has applied a Creative Commons Attribution (CC BY) licence to any Author Accepted Manuscript version arising from this submission.
% RMS acknowledges support through Australian Research Council Future Fellowship FT190100155 and Discovery Project DP220102305. 
% A.C.G. and the Fong Group at Northwestern acknowledges support by the National Science Foundation under grant Nos. AST-1909358, AST-2308182 and CAREER grant No. AST-2047919. 
% J.X.P., A.C.G. acknowledge support from NSF grants AST-1911140, AST-1910471 and AST-2206490 as members of the Fast and Fortunate for FRB Follow-up team. 
% ATD acknowledges support through Australian Research Council Discovery Project DP22010230.

\paragraph{Competing Interests}
None.

\paragraph{Data Availability Statement}
The code and data used to produce our results can be found at \href{https://github.com/FRBs/zdm}{https://github.com/FRBs/zdm}.

\printendnotes

% \printbibliography 
\bibliography{References.bib}

\end{document}